\documentclass{article}
\usepackage{spconf}
\usepackage{KalmanNet,adjustbox}
\usepackage[bookmarks,colorlinks]{hyperref}
\usepackage{cite}
\usepackage[table, dvipsnames]{xcolor}
\usepackage{subcaption}

\newcommand{\myVec}[1]{{\boldsymbol{#1}}}

\title{Adaptive KalmanNet: Data-Driven Kalman Filter with Fast Adaptation}
\name{Xiaoyong Ni, Guy Revach, and Nir Shlezinger
\thanks{
X. Ni and G. Revach are with ISI, D-ITET, ETH Zürich, 
Switzerland (e-mail: xiaoni@student.ethz.ch, grevach@ethz.ch). 
N. Shlezinger is with the School of ECE, Ben-Gurion University of the Negev, Beer Sheva, Israel (e-mail: nirshl@bgu.ac.il). 
We thank H-A. Loeliger for helpful discussions.}}
\address{\vspace{-25mm}}
\begin{document}
\maketitle
%
%
\begin{abstract}
Combining the classical \ac{kf} with a \ac{dnn} enables tracking in partially known \ac{ss} models. A major limitation of current \ac{dnn}-aided designs stems from the need to train them to filter data originating from a specific distribution and underlying \ac{ss} model. Consequently, changes in the model parameters may require lengthy retraining. While the \ac{kf} adapts through parameter tuning, the black-box nature of \acp{dnn} makes identifying tunable components difficult. Hence, we propose \ac{aknet}, a \ac{dnn}-aided \ac{kf} that can adapt to changes in the \ac{ss} model without retraining. Inspired by recent advances in \acl{llm} fine-tuning paradigms, \ac{aknet} uses a compact hypernetwork to generate \acl{cm} weights. Numerical evaluation shows that \ac{aknet} provides consistent state estimation performance across a continuous range of noise distributions, even when trained using data from limited noise settings.
\end{abstract}
%
%
\acresetall 
%
\vspace{-0.2cm}
\section{Introduction}\label{sec:intro}
\vspace{-0.15cm}
Estimating the hidden state of a dynamic system from noisy observations is crucial in a wide range of applications~\cite{durbin2012time}.  
Traditional \ac{mb} methods, such as the \ac{kf} \cite{kalman1960new}, leverage mathematical parametric representations in the form of \ac{ss} models that describe the underlying dynamics. The reliance of the \ac{kf} and its variants on knowledge of the \ac{ss} model implies that they are inherently adaptive, in the sense that changes in the model parameters are naturally incorporated into its operation. However, they are also sensitive to mismatches in the \ac{ss} model, and are most suitable for models with Gaussian noises~\cite[Ch. 10]{durbin2012time}. 

Over the recent years, \ac{dnn}-based filters have emerged as  \ac{dd} alternatives to \ac{mb} filters. Highly parameterized \acp{dnn} can be trained \acl{e2e} using massive datasets for filtering without relying on the \ac{ss} model~\cite{becker2019recurrent}. Alternatively, one can fuse principled statistical models with a \ac{dd} process via \acl{mb} \acl{dl}~\cite{shlezinger2020model,shlezinger2023model}, where the flow of the \ac{kf} is preserved based on some of the SS model parameters and augmented with compact \acp{dnn}~\cite{revach2022kalmannet,ghosh2023danse,choi2023split,revach2021rtsnet}. While hybrid \ac{mb}/\ac{dd} designs offer greater flexibility than their counterparts and support adaptation through compact \acp{dnn}~\cite{KalmannetICAS21} (trainable with smaller datasets) and unsupervised learning~\cite{revach2022unsupervised,ghosh2023danse}, they lack the inherent adaptivity of \ac{mb} designs via mere parameter tuning. Adjusting a  \ac{dd} system to distribution shifts typically involves time-consuming and computationally intensive retraining~\cite{raviv2023adaptive}.

In this work, we introduce {\em \ac{aknet}}, an adaptive \ac{mb}/\ac{dd} filter that is trained with data to cope with a model mismatch, and can rapidly adapt to changes in the \ac{ss} model without retraining. Our \ac{aknet} extends \acl{kn}\cite{revach2022kalmannet} by adapting its mapping based on a context information parameter coined {\em \ac{sow}}. This \ac{sow} is used as an input to a hypernetwork\cite{ha2016hypernetworks,galanti2020modularity}, which fine-tunes  \acl{kn}'s \ac{dnn} to adapt to different contexts. When tracking in face of partially-known non-Gaussian \ac{ss} models, the \ac{sow} serves as an indicator for the variances of the noise signals.

Unlike previously proposed hypernetworks~\cite{goutay2020deep,liu2022learning,welling2021hkf}, \ac{aknet} is tailored for a compact implementation, and it draws inspiration from recent \ac{cm} techniques in \acp{llm} — used for fine-tuning general \acp{llm} to specialized tasks~\cite{ding2023parameter}. Our approach achieves a significant reduction in trainable parameters, outperforming both alternative hypernetworks~\cite{ha2016hypernetworks} and ensemble (filter-bank) architectures~\cite{khodarahmi2023review}. To facilitate "learning to filter" across different \ac{ss} models, we propose a dedicated two-stage training method. In numerical evaluations, \ac{aknet} {consistently} estimates states across a continuous range of unseen distributions with varied noise variances, even when trained on limited data from a discrete set of distributions. Furthermore, it successfully tracks rapidly changing distributions and shows robustness against errors in \ac{sow} estimation.
%
%

The remainder of this paper is structured as follows: Section~\ref{sec:Poblem Formulation} formulates the problem of state estimation with fast adaptation; Section~\ref{sec:Adaptive KalmanNet} details \ac{aknet}; Section 4 presents our numerical study; and Section~\ref{sec:Conclusions} concludes the paper.
%

%
%
\vspace{-0.35cm}
\section{PROBLEM FORMULATION}\label{sec:Poblem Formulation}
\vspace{-0.15cm}
%
%
\subsection{State Estimation}
\vspace{-0.1cm}
We consider dynamical systems characterized by a \ac{ss} model in discrete-time. We focus on linear models with unknown time-varying noise signals, which (possibly) follow non-Gaussian distributions
\begin{subequations}\label{eq:ssmodel}
\begin{align}\label{eqn:stateEvolution}
\gvec{x}_{t}&= 
\gvec{F}\cdot \gvec{x}_{t-1}+\gvec{e}_{t},& 
{\rm Var}(\gvec{e}_t) = \gvec{Q}_t, \quad
&\gvec{x}_{t}\in\greal^m,\\
\label{eqn:stateObservation}
\gvec{y}_{t}&=
\gvec{H}\cdot \gvec{x}_{t}+\gvec{v}_{t},
& 
{\rm Var}(\gvec{v}_t) = \gvec{R}_t, \quad & 
\gvec{y}_{t}\in\greal^n.    
\end{align}
%
\end{subequations}
Here, $\gvec{x}_{t}$ is the latent state vector  at time $t$, which is evolved by a state evolution matrix $\gvec{F}$, and by an additive zero-mean process noise $\gvec{e}_t$. The vector $\gvec{y}_{t}$ represents the observations at time $t$, which is generated from the latent state by a linear mapping $\gvec{H}$, and corrupted by an additive  zero-mean noise $\gvec{v}_t$. 

%
\vspace{-0.1cm}
\subsection{Preliminaries}
\label{ssec:Kalman}
\vspace{-0.1cm}
Given knowledge of $\gvec{Q}_t$ and $\gvec{R}_t$, \ac{kf} is \acl{mse} optimal if $\gvec{e}_t, \gvec{v}_t$ both follow Gaussian distributions. 
This is achieved by first predicting the next state and observations based on the previous estimate $\hat{\gvec{x}}_{t-1}$ via 
\begin{equation}
\label{eqn:predict}
    \hat{\gvec{x}}_{t|t-1} = \gvec{F}\cdot \hat{\gvec{x}}_{t-1}, \quad \hat{\gvec{y}}_{t|t-1} = \gvec{H}\cdot  \hat{\gvec{x}}_{t|t-1}.
\end{equation}
The next estimate $\hat{\gvec{x}}_{t}$ is obtained using the observed $\gvec{y}_t $ as  follows
\begin{equation}
    \label{eqn:update}
     \hat{\gvec{x}}_{t}  =  \hat{\gvec{x}}_{t|t-1}  + \Kgain_t\cdot  (\gvec{y}_t - \hat{\gvec{y}}_{t|t-1}),
\end{equation}
where $\Kgain_t $ is the \ac{kg} computed by tracking the second order moments, using  $\gvec{Q}_t$ and $\gvec{R}_t$. 
For unknown $\gvec{Q}_t$ and $\gvec{R}_t$, a range of adaptive \ac{kf}  methods \cite{zhang2020identification} have been proposed, with the main idea of adding a noise estimator on top of \ac{kf}. These \ac{mb} approaches manage noise shifts conveniently through tuning the parameters $\gvec{Q}_{t}$ and $\gvec{R}_{t}$ that are fed into the \ac{kf}.

The \ac{dnn}-aided \acl{kn}~\cite{revach2022kalmannet} is trained to produce state estimates in a discriminative manner~\cite{shlezinger2022discriminative}. It can successfully learn from data to cope with non-Gaussian distributions. This is achieved by preserving the operation of the \ac{kf} in \eqref{eqn:predict}-\eqref{eqn:update}, while computing the \ac{kg} $\Kgain_t $ using a \ac{dnn} with parameters $\myVec{\theta}$, comprised of a \ac{rnn} with preceding and subsequent \ac{fc} layers, denoted  $\Kgain_t(\myVec{\theta})$, applied to features extracted from $\gvec{y}_t $  and $\hat{\gvec{x}}_t$.


%
\vspace{-0.1cm}
\subsection{Fast Adaptation Problem}
\vspace{-0.1cm}


 \ac{dnn}-aided filters such as \acl{kn} can cope with non-Gaussian noises and possible mismatches in $\gvec{F}$ and $\gvec{H}$; however, they are typically trained for a specific \ac{ss} model.  The parameters $\myVec{\theta}$ of \acl{kn} are trained using  data corresponding to (at most) a limited set of distributions.
 In our time-varying setting, a typical \ac{kf} implementation would involve an additional estimator for recovering $\gvec{Q}_t$ and $\gvec{R}_t$ which can be substituted into its computation of  $\Kgain_t$. This ability is not supported by \ac{dnn}-aided filters such as \acl{kn}, even when one has access during run-time to instantaneous estimates of $\gvec{Q}_t$ and $\gvec{R}_t$. Thus, we wish to extend \acl{kn} to reliably track in \ac{ss} models with time-varying noise statistics. 

In particular, we do not assume full knowledge of matrices $\gvec{Q}_t$ and $\gvec{R}_t$, but rather only to the scalar ${\rm SoW}_t$ indicating the rough scaling ratio between process noise and observation noise, given by
\begin{equation}\label{eqn:opt sow}
    {\rm SoW}_t = \frac{n \cdot {\rm Tr}(\gvec{Q}_t)}{m \cdot {\rm Tr}(\gvec{R}_t)}.
\end{equation}
This formulation of the \ac{sow} is selected due to it being sufficient statistics for computing the \ac{kg} in linear Gaussian \ac{ss} models with scaled identity noise variance matrices~\cite{61006}.  We assume that the system has access to \eqref{eqn:opt sow}, possibly provided by some external estimator, noting that estimating this scalar quantity is expected to be notably simpler compared to estimating the matrices  $\gvec{Q}_t$ and $\gvec{R}_t$.

The training dataset comprises $n_t$ length $T$ state trajectories alongside their corresponding \acp{sow}, i.e., 
\begin{equation}
\label{eqn:dataset}
\mathcal{D} = \set{\set{\gvec{x}_t^{(i)}, \gvec{y}_t^{(i)}, {\rm SoW}_t^{(i)}}_{t=1}^{T}}_{i=1}^{n_t}. 
\end{equation}
The \acp{sow} in \eqref{eqn:dataset} may cover only a few discrete points of noise settings. During inference, the system is required to be able to track over a continuous range of noise settings.  

%
\vspace{-0.2cm}
\section{Adaptive KalmanNet}\label{sec:Adaptive KalmanNet}
\vspace{-0.15cm}


%
\subsection{Architecture}\label{subsec:Arch}
\vspace{-0.1cm}
%
%
\begin{figure}
\centering
\includegraphics[width=0.5\textwidth]{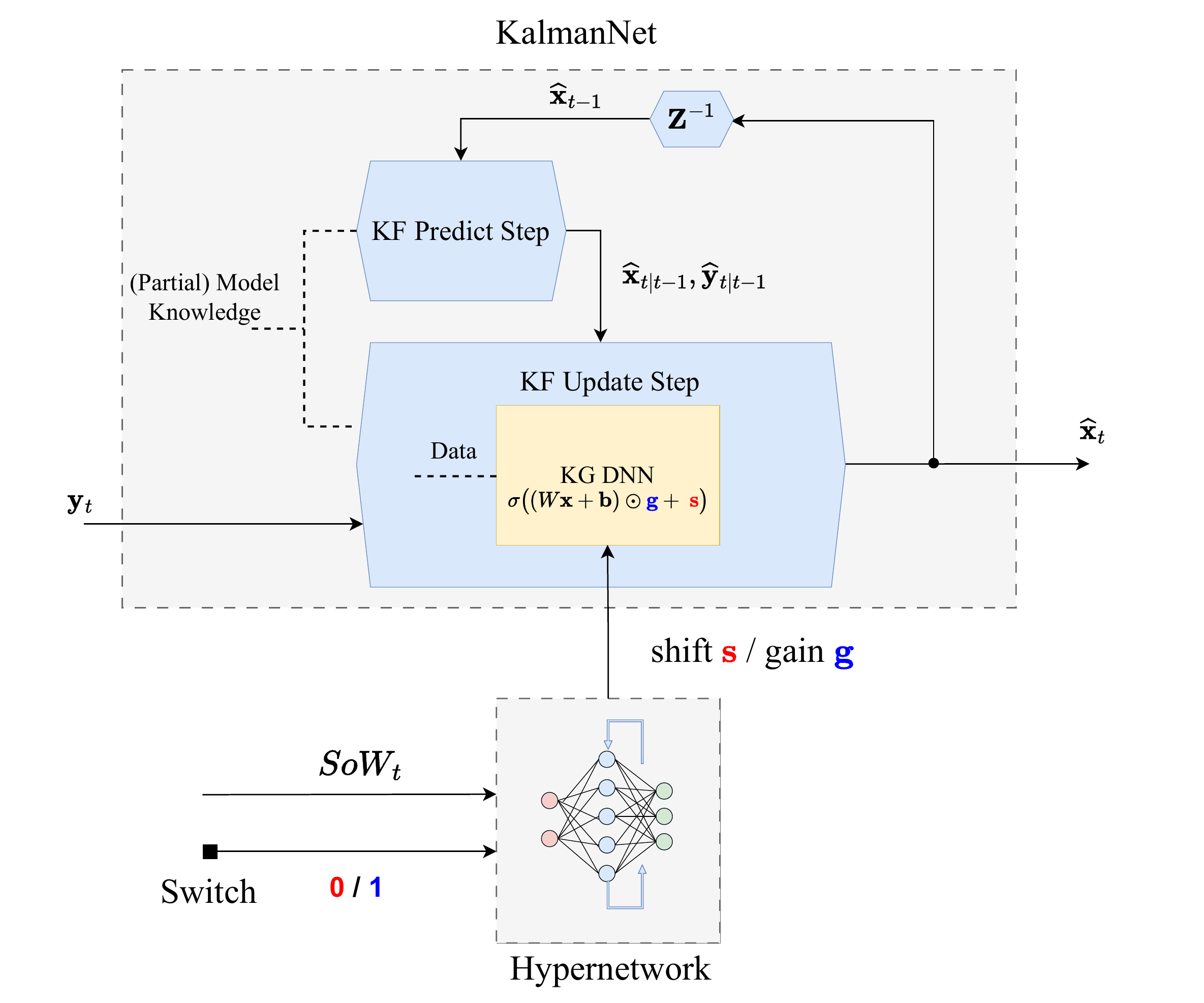}
\caption{Overall architecture of AKNet}
\label{fig:overall arch}
\vspace{-0.2cm}
\end{figure}
%

\ac{aknet}  is an extension of \acl{kn}, designed to handle time-varying noise distributions. 
Given the context ${\rm SoW}_t$, a compact hypernetwork with parameters $\myVec{\psi}$ generates \ac{cm} weights. These \ac{cm} weights then fine-tune the \ac{kg} \ac{dnn} parameters, such that the \ac{kg} computed by \acl{kn} becomes $\Kgain_t(\myVec{\theta},{\rm SoW}_t;\myVec{\psi})$.



\textbf{Hypernetwork:} In general, hypernetworks utilize additional \acp{dnn} to generate the parameters of another \ac{dnn}~\cite{ha2016hypernetworks}. This allows the \ac{dnn} mapping to be influenced by additional features. Hypernetworks are typically very highly parameterized, with many output neurons dictated by the number of \ac{dnn} parameters, making them computationally expensive and challenging to train.
To exploit the ability of hypernetworks to set the \ac{kg} \ac{dnn} mapping based on ${\rm SoW}_t$ without dramatically increasing parameterization, we employ \ac{cm}~\cite{ding2023parameter}. Instead of generating all the parameters of \ac{kg} \ac{dnn}, \ac{cm} fine-tunes the original parameters.
The hypernetwork with parameters $\myVec{\psi}$ induces a mapping  $d_{\myVec{\psi}}$  with \ac{cm} weights as output: gain $\mathbf{g}$
and shift $\mathbf{s}$, whereas gain represents a multiplicative modulation and shift represents an additive modulation, respectively. To be parameter efficient, we reuse the same hypernetwork to generate $\mathbf{g}$ and $\mathbf{s}$ through an additional  input ${\rm switch} \in \{0,1\}$, such that
\begin{equation}
    \mathbf{g} = d_{\myVec{\psi}}({\rm SoW}_t, 1), \quad \mathbf{s} = d_{\myVec{\psi}}({\rm SoW}_t, 0).
    \label{eqn:CMGain}
\end{equation}

\textbf{\acl{cm}:} \ac{cm} performs on the level of neuronal computation. It is  applied in \ac{aknet} to both \ac{fc} layers and \ac{rnn} layers of \acl{kn}. 
To formulate the operation of \ac{cm}, we consider a generic form of linear computation with weights $\gvec{W}$ and bias $\gvec{b}$ applied with a with nonlinear activation function $\sigma(\cdot)$, i.e., $\sigma(\gvec{W} \mathbf{x} + \mathbf{b})$. This form describes both \ac{fc} layers and the gate computations in \ac{rnn}. 
With \ac{cm} applied, the operation of such a layer becomes
\begin{equation}\label{eqn:CM gain and shift}
\sigma \big( (\gvec{W} \mathbf{x} + \mathbf{b}) \odot \mathbf{g} + \
    \mathbf{s} \big),
\end{equation}
where $\odot$ represents element-wise multiplication, and $\gvec{g},\gvec{s}$ are the gains and shifts obtained from the neurons of the hypernetwork corresponding to the current layer via \eqref{eqn:CMGain}. The resulting architecture is illustrated in Fig.~\ref{fig:overall arch}.

 
%
%
\vspace{-0.1cm}
\subsection{Training}
\vspace{-0.1cm}
 
The training of \ac{aknet}  is based on the $\ell_2$ loss, which for a dataset $\mathcal{D}$ is given by (omitting weights regularization) 
\begin{equation}
    \label{eqn:Loss}
    \mathcal{L}_{\mathcal{D}}(\myVec{\theta},\myVec{\psi}) = 
    \frac{1}{|\mathcal{D}| T}\sum_{i=1}^{|\mathcal{D}|}\sum_{t=1}^T \| \gvec{x}_t^{(i)} - \hat{\gvec{x}}_t( \gvec{y}_t^{(i)}; \myVec{\theta},\myVec{\psi})\|^2.
\end{equation}
In \eqref{eqn:Loss}, $\hat{\gvec{x}}_t( \gvec{y}_t; \myVec{\theta},\myVec{\psi})$ is the estimation of $\gvec{x}_t$ produced from $\gvec{y}_t$ by \ac{aknet} with parameters $\myVec{\theta},\myVec{\psi}$. 

We train \ac{aknet} in two stages. We first train only \acl{kn}. To that aim, we fix the \ac{cm} layer not to affect the \ac{kg} computation, i.e., to output unit gains and zero shift, and extract a subset $\tilde{\mathcal{D}}\subset \mathcal{D}$ where all trajectories have a relatively stationary and similar noise distributions. We then train solely $\myVec{\theta}$ based on the loss $\mathcal{L}_{\tilde{\mathcal{D}}}$.
In the second stage, we freeze $\myVec{\theta}$, and train only the hypernetwork parameters $\myVec{\psi}$. Here, we use the entire dataset $\mathcal{D}$, which encompasses non-stationary noise distributions, and train $\myVec{\psi}$ with the fixed $\myVec{\theta}$ based on  $\mathcal{L}_{{\mathcal{D}}}$.

    
    


    

%
\vspace{-0.1cm}
\subsection{Discussion}\label{subsec:discussion}
\vspace{-0.1cm}
\ac{aknet} allows \ac{dnn}-aided tracking to be carried out in \ac{ss} models with time-varying noise statistics without requiring retraining. It is based on \acl{kn} due to its preservation of the majority of the \ac{mb} attributes intrinsic to the \ac{kf}. 
As \ac{kg} effectively encodes the information regarding the noise statistics, \ac{aknet} enables adaptation by augmenting it with a dedicated hypernetwork. 
It uses a hypernetwork based on \ac{cm} due to its parameter efficiency and quick adaptation. The former is illustrated in Table~\ref{tab:Parameter efficient CM}, reporting the number of trainable parameters of \ac{aknet} used for different \ac{ss} model sizes. We observe in Table~\ref{tab:Parameter efficient CM} that the number of \ac{cm} weights is much smaller than that of \acl{kn}, showcasing superior parameter efficiency over employing ensemble (filter-bank) designs. The trained hypernetwork, which maps ${\rm SoW}_t$ into \ac{cm} weights, facilitates fast adaptation in  online inference. 
The overall  approach addresses a frequent challenge in \ac{dd} methods: the ambiguous task of determining which parameters require tuning for a specific shift. The rationale used in \ac{aknet} can be extended to alternative hybrid \ac{mb}/\ac{dd} systems employed in time-varying conditions.

Our problem formulation considers the \ac{sow} as being externally provided, while in practice, it should be estimated. There are various ways to estimate ${\rm SoW}_t$. This noise estimator design is mainly a tradeoff between robustness and inference speed. For example, \acl{em} \cite{dauwels2005expectation}  algorithm is more robust since its convergence can be guaranteed, while  correlation-based methods \cite{mehra1970identification, akhlaghi2017adaptive} using one-step estimation can be much faster while less guaranteed in terms of performance. In our case, since ${\rm SoW}_t$ is only a scalar, a simple estimation method is grid search with unsupervised loss as criteria. Alternatively, it can be based on a machine learning estimator that is jointly trained alongside the hypernetwork. 
In our numerical study in Section~\ref{sec:NumEval}, we show that \ac{aknet} is robust to errors in the \ac{sow}, and leave its study   with \ac{sow} estimation for future work.

\begin{table}
\centering
\begin{adjustbox}{width=\columnwidth} 
\begin{tabular}{|c|c|c|c|c|}
\hline $m\times n$& \acl{kn}  & Hypernet \cite{ha2016hypernetworks} & {\ac{cm}} & $p$  Ensemble\\
\hline {$2\times2$} & 10k & 6k & 1k & 
 $p \cdot 10k$\\
\hline {$2\times2$} & 22k & 90k & 1.5k & $p \cdot 20k$\\
\hline {$10\times10$} & 330k & 100k & 6k & $p \cdot 300k$\\
\hline 
\end{tabular}
\end{adjustbox}
\caption{System size vs. number of parameters}
\label{tab:Parameter efficient CM}
\vspace{-0.5cm}
\end{table}
%


\vspace{-0.2cm}
\section{Numerical Evaluations}\label{sec:NumEval}
\vspace{-0.15cm}
In this section we provide a numerical evaluation of \ac{aknet}. We consider three \ac{ss} models with different noise distributions\footnote{The source code can be found online at \url{https://github.com/KalmanNet/Adaptive-KNet-ICASSP24}.}: $(i)$ a  Gaussian setting, showcasing the ability of \ac{aknet} to achieve the optimal \acl{mse} of the \ac{mb} \ac{kf};
$(ii)$ a non-Gaussian setting, demonstrating \ac{aknet}'s gains in coping with non-Gaussian noises;
and $(iii)$ a setting with noisy \acp{sow}, studying the robustness of \ac{aknet} to errors in this feature. 
Unless stated otherwise,  we set $\gvec{Q}_t = q_t^2\cdot\gvec{Q}_0$ and $\gvec{R}_t = r_t^2\cdot\gvec{R}_0$ with $m=n$, such that ${\rm SoW}_t=\frac{q_t^2}{r_t^2}$. 
The pseudo-stationary dataset $\tilde{\mathcal{D}}$ used to train $\myVec{\theta}$ has 100 trajectories with noise setting $\gvec{Q}_0$, $\gvec{R}_0$,  while $\mathcal{D}$ used to the train the hypernetwork $\myVec{\psi}$ has  400 trajectories. 
%
%

\begin{figure}
\begin{subfigure}{.25\textwidth} 
\includegraphics[width=1\linewidth]{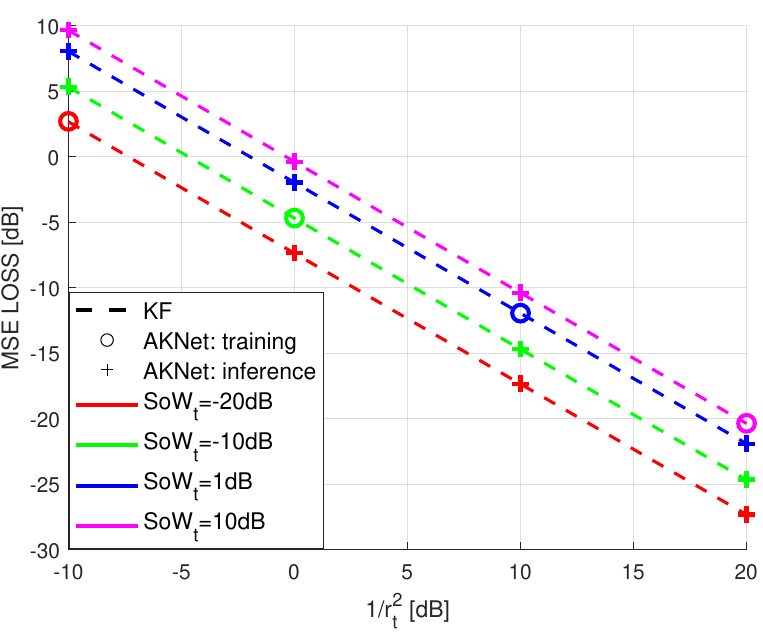}
\label{fig:train sow}
\end{subfigure}%
\begin{subfigure}{.25\textwidth} 
\includegraphics[width=1\linewidth]{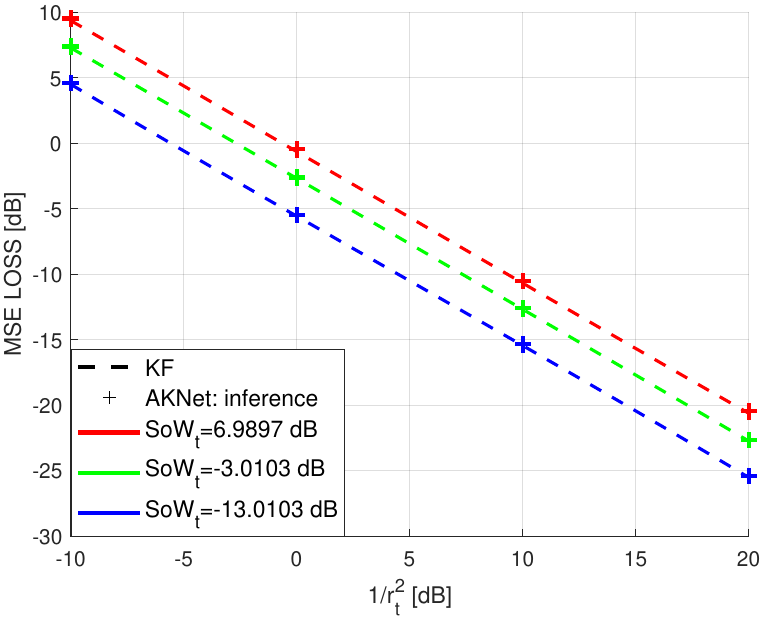}
\label{fig:test sow}
\end{subfigure}
\caption{Linear Gaussian system: \ac{aknet} trained on four discrete value pairs of $(q_t^2, r_t^2)$ (marked as circles) and tested on shifts preserving the same ratio (left) and unseen ratios (right).}
\label{fig:base+cm train and test on diff sow}
\vspace{-0.2cm}
\end{figure}

{\bf Gaussian Noise:}
First, we evaluate \ac{aknet} on a linear Gaussian \ac{ss} model. 
We randomly set $\gvec{Q}_0$ and $ \gvec{R}_0$, only requiring them to be positive definite (no need to be diagonal).
The data set $\mathcal{D}$ contains only four different noise variance pairs. 
In Fig.~\ref{fig:base+cm train and test on diff sow}, the dashed lines represent the performances of the \ac{kf} on these datasets, serving as an optimal baseline given the linear Gaussian setting. The figure reveals that \ac{aknet} not only coincides with the \ac{kf} for the four \ac{sow}  observed during training, but also generalizes to unseen distributions. This generalization includes both \ac{ss} models with the same ${\rm SoW}_t$ ratio but with differing scaling values $q_t^2, r_t^2$, as well as \ac{ss} models with ${\rm SoW}_t$ ratios that are unseen during training.
This demonstrates that \ac{aknet} only needs a small amount of training data on a limited number of settings in order for it to handle a wide varying range of noise statistics. 

\begin{figure}
\begin{subfigure}{.25\textwidth} 
\includegraphics[width=1\linewidth]{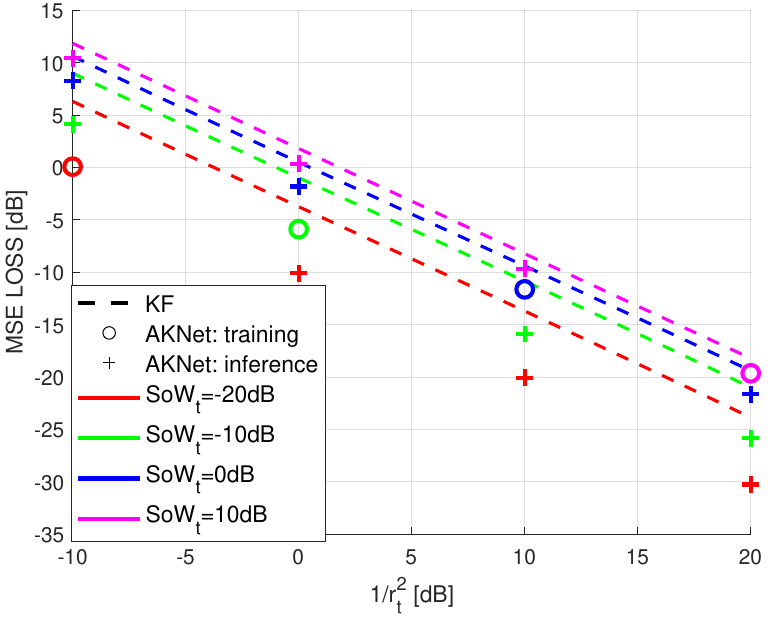}
\label{fig:exp train sow}
\end{subfigure}%
\begin{subfigure}{.25\textwidth} 
\includegraphics[width=1\linewidth]{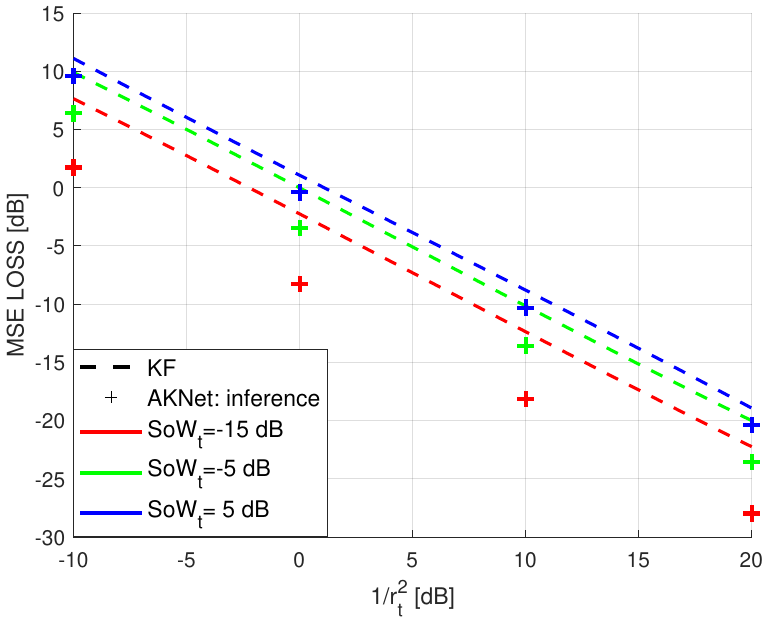}
\label{fig:exp test sow}
\end{subfigure}
\caption{Linear non-Gaussian system: \ac{aknet} trained on four  value pairs of $(q_t^2, r_t^2)$ (marked as circles) and tested on shifts preserving the same ratio (left) and unseen ratios (right).}
\label{fig:base+cm for exp}
\vspace{-0.2cm}
\end{figure}

%
%
{\bf Non-Gaussian Noise:}
We use exponentially distributed noise signals that are spatially uncorrelated, i.e., $\gvec{Q}_0$ and $\gvec{R}_0$  are scaled identity matrices. 
We again train \ac{aknet} using merely four different distribution pairs, and test it on \ac{ss} models with different distributions that either preserve \acp{sow} observed in training as well as \acp{sow} seen only in inference. 
The results, reported in Fig.~\ref{fig:base+cm for exp}, highlight that the two forms of generalization capabilities are still kept even for the non-Gaussian case. Furthermore, \ac{aknet} can significantly outperform classic \ac{kf}, due to the suitability of \acl{kn} in handling non-Gaussian distributions.



\begin{figure}
\centering
\includegraphics[width=0.39\textwidth]{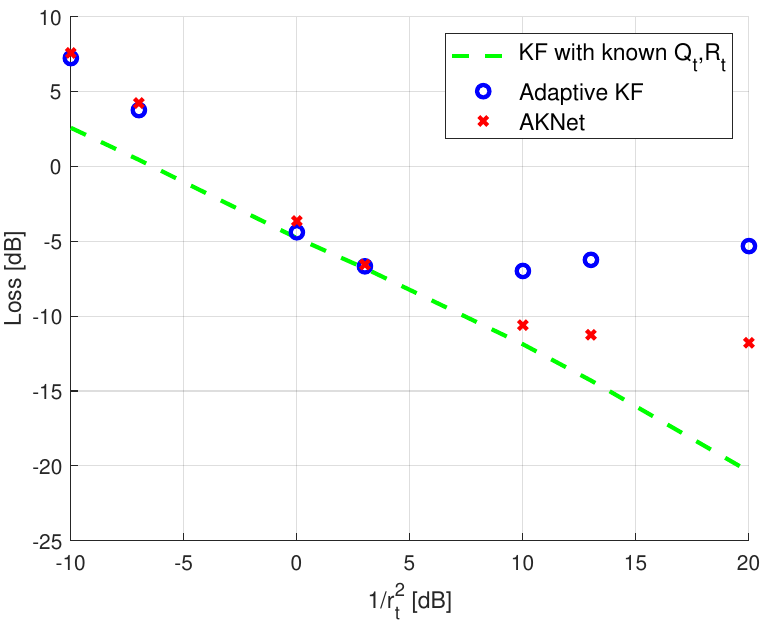}
\caption{Time-varying SoW}
\label{fig:tracking}
\vspace{-0.3cm}
\end{figure}

%
%
{\bf Noisy \acp{sow}:}
So far, we have shown that \ac{aknet} incorporates a continuous shifting range of noise distributions in its compact network model. We next study its ability to cope with noisy \ac{sow}, arising from estimation errors encountered during online inference with time-varying noise distributions. 
We again consider a Gaussian \ac{ss} model, and train \ac{aknet} with the same data as that used in Fig.~\ref{fig:base+cm train and test on diff sow}. During inference, the time-variations simulate abrupt jumps. Particularly, we assume that the scaling parameters $q_t^2, r_t^2$  jump to different values in each timestep $t$. The simulation results shown in Fig.~\ref{fig:tracking} are generated with ground truth scaling parameters $(q^2_{t-1} = 1, 
r^2_{t-1} = 1)$ at previous timestep 
and jump to 
$(q^2_t = 0.1,
r^2_t = [0.01, 0.05, 0.1, 0.5, 1, 5, 10])$.
For fair comparison, \ac{aknet} uses the same correlation-based noise estimator \cite{akhlaghi2017adaptive} as the adaptive \ac{kf}. 

In Fig.~\ref{fig:tracking}, both \ac{kf} and \ac{aknet} remain optimal when the jumping step of $r_t^2$ is not large. However, as the jumping step increases, especially in low observation noise scenarios, adaptive \ac{kf} produces much worse state estimation than \ac{aknet}. 
In summary, \ac{aknet} can keep track of jumping noise distributions, approaching optimal state estimation provided the jumping step remains within a certain limit. If it is outside the limit, \ac{aknet} can still do better than classic \ac{kf} even in this linear Gaussian setting. The correlation-based noise estimator we choose values inference speed over accuracy, showing the robustness of \ac{aknet} to noise estimation errors.

%

%
\vspace{-0.3cm}
\section{Conclusions}\label{sec:Conclusions}
\vspace{-0.15cm}
We have presented \ac{aknet}, a \ac{dnn}-aided \ac{kf} capable of handling varying noise statistics using a single parameter tuning. \ac{aknet} utilizes a compact hypernetwork to generate \ac{cm} weights and employs a two-stage training process. Numerical evaluation shows \ac{aknet}’s adaptability to varying
noise distributions during state estimation and robustness in online tracking, with noise estimation errors. Although primarily tailored for state estimation, the fundamental principles could potentially be adapted for other DD signal processing systems in dynamic scenarios.
%
%
\bibliographystyle{IEEEtran}
\bibliography{IEEEabrv,KalmanNet}
\end{document}